\documentclass[12pt]{article}
\usepackage{epsfig}

\renewcommand{\theequation}{\arabic{section}.\arabic{equation}}

\def\beq{\begin{eqnarray}}
\def\eeq{\end{eqnarray}}
\def\bea{\begin{eqnarray}}
\def\eea{\end{eqnarray}}

\def\tev{\, {\rm TeV}}
\def\gev{\, {\rm GeV}}

\newcommand{\gsim}{\lower.7ex\hbox{$\;\stackrel{\textstyle>}{\sim}\;$}}
\newcommand{\lsim}{\lower.7ex\hbox{$\;\stackrel{\textstyle<}{\sim}\;$}}

\def\Ni{\tilde \chi^0_i}
\def\Nj{\tilde \chi^0_j}
\def\Ci{\widetilde \chi^+_i}
\def\Cj{\widetilde \chi^+_j}
\def\lnbar{\overline{\rm ln}}

\def\stilde{\widetilde}

\newcommand{\newc}{\newcommand}
\newc{\Nc}{N_{c}}
\newc{\CG}{C_G}
\newc{\gp}{g'}
\newc{\stopi}{\stilde t_i}
\newc{\sboti}{\stilde b_i}
\newc{\staui}{\stilde \tau_i}
\newc{\stopj}{\stilde t_j}
\newc{\sbotj}{\stilde b_j}
\newc{\stauj}{\stilde \tau_j}
\newc{\stopI}{\stilde t_1}
\newc{\stopII}{\stilde t_2}
\newc{\sbotI}{\stilde b_1}
\newc{\sbotII}{\stilde b_2}
\newc{\stauI}{\stilde \tau_1}
\newc{\stauII}{\stilde \tau_2}
\newc{\sstop}{s_{t}}
\newc{\cstop}{c_{t}}
\newc{\ssbot}{s_{b}}
\newc{\csbot}{c_{b}}
\newc{\sstau}{s_{\tau}}
\newc{\cstau}{c_{\tau}}
\newc{\Sstop}{s_{2t}}
\newc{\Cstop}{c_{2t}}
\newc{\Ssbot}{s_{2b}}
\newc{\Csbot}{c_{2b}}
\newc{\Sstau}{s_{2\tau}}
\newc{\Cstau}{c_{2\tau}}
\newc{\salpha}{s_\alpha}
\newc{\calpha}{c_\alpha}
\newc{\Calpha}{c_{2\alpha}}
\newc{\Salpha}{s_{2\alpha}}
\newc{\sbetapm}{s_{\beta_\pm}}
\newc{\cbetapm}{c_{\beta_\pm}}
\newc{\Sbetapm}{s_{2 \beta_\pm}}
\newc{\Cbetapm}{c_{2 \beta_\pm}}
\newc{\sbetaO}{s_{\beta_0}}
\newc{\cbetaO}{c_{\beta_0}}
\newc{\SbetaO}{s_{2 \beta_0}}
\newc{\CbetaO}{c_{2 \beta_0}}
\newc{\vu}{v_u}
\newc{\vd}{v_d}
\newc{\seL}{\stilde e_L}
\newc{\smuL}{\stilde \mu_L}
\newc{\seR}{\stilde e_R}
\newc{\smuR}{\stilde \mu_R}
\newc{\suL}{\stilde u_L}
\newc{\sdL}{\stilde d_L}
\newc{\suR}{\stilde u_R}
\newc{\sdR}{\stilde d_R}
\newc{\scL}{\stilde c_L}
\newc{\ssL}{\stilde s_L}
\newc{\scR}{\stilde c_R}
\newc{\ssR}{\stilde s_R}
\newc{\snue}{\stilde \nu_e}
\newc{\snumu}{\stilde \nu_\mu}
\newc{\snutau}{\stilde \nu_\tau}
\newc{\Gpm}{G^\pm}
\newc{\Hpm}{H^\pm}
\newc{\FFbS}{\overline{FF}S}
\newc{\FFbV}{\overline{FF}V}
\newc{\FSS}{F_{SS}}
\newc{\FSSS}{F_{SSS}}
\newc{\FFFS}{F_{FFS}}
\newc{\FFFbS}{F_{\overline{FF}S}}
\newc{\FSSV}{F_{SSV}}
\newc{\FVS}{F_{VS}}
\newc{\FVVS}{F_{VVS}}
\newc{\FFFV}{F_{FFV}}
\newc{\FFFbV}{F_{\overline{FF}V}}
\newc{\Fgauge}{F_{\rm gauge}}
\newc{\DRbarprime}{$\overline{\rm DR}'$ }
\newc{\DRbar}{$\overline{\rm DR}$ }
\newc{\MSbar}{$\overline{\rm MS}$ }
\newc{\Yu}{{\bf Y}_u}
\newc{\Yd}{{\bf Y}_d}
\newc{\Ye}{{\bf Y}_e}
\newc{\Au}{{\bf a}_u}
\newc{\Ad}{{\bf a}_d}
\newc{\Ae}{{\bf a}_e}
\newc{\bm}{{\bf m}}

\newc{\rwino}{r_{\tilde W}}
\newc{\rmu}{r_{\tilde H}}
\newc{\ra}{r_A}

\begin{document}

\setlength{\baselineskip}{0.25in}


\begin{titlepage}
\noindent
\begin{flushright}
MSUHEP-041221 \\
MCTP-04-74 \\
\end{flushright}
\vspace{1cm}

\begin{center}
  \begin{Large}
    \begin{bf}
Virtual effects of light gauginos and higgsinos: \\
a precision electroweak analysis of split supersymmetry

    \end{bf}
  \end{Large}
\end{center}
\vspace{0.2cm}
\begin{center}
\begin{large}
Stephen P. Martin$^a$, Kazuhiro Tobe$^b$, James D. Wells$^{c}$ \\
\end{large}
  \vspace{0.3cm}
  \begin{it}
${}^{(a)}$Physics Department, Northern Illinois University, DeKalb, IL 60115\\
~~{\rm and} Fermi National Accelerator Laboratory, PO Box 500, Batavia, IL 60510 \\
\vspace{0.1cm}
${}^{(b)}$Department of Physics and Astronomy \\ 
Michigan State University, East Lansing, MI 48824 \\
\vspace{0.1cm}
${}^{(c)}$Michigan Center for Theoretical Physics (MCTP) \\
        ~~University of Michigan, Ann Arbor, MI 48109-1120, USA \\
\vspace{0.1cm}
\end{it}

\end{center}

\begin{abstract}

We compute corrections to precision electroweak observables
in supersymmetry in the limit that scalar superpartners
are very massive and decoupled.  This leaves charginos and
neutralinos and a Standard Model-like Higgs boson as the 
only states with unknown mass substantially affecting the
analysis.  We give complete formulas for the chargino and neutralino
contributions, derive simple analytic results for the 
pure gaugino and higgsino cases, and study the general case.  
We find that in all circumstances,
the precision electroweak fit improves 
when the charginos and neutralinos are near the current direct
limits.  Larger higgsino and gaugino masses worsen the fit as the 
theory predictions asymptotically approach those of the Standard Model.  
Since the Standard
Model is considered by most to be an adequate fit to the precision
electroweak data, an important corollary to our analysis is that
all regions of parameter space allowed by direct
collider constraints are also allowed by precision electroweak
constraints in split supersymmetry.

\end{abstract}

\vspace{1cm}

\begin{flushleft}
hep-ph/0412424 \\
December 2004
\end{flushleft}

\end{titlepage}

\setcounter{footnote}{1}
\setcounter{page}{2}
\setcounter{figure}{0}
\setcounter{table}{0}


\section{Introduction}

Ordinary intuition about finetuning and naturalness applied to
supersymmetry implies the existence of light (TeV or less) scalar
superpartners, gauginos and higgsinos.  
Supersymmetry has several other good reasons for its existence
beyond its ability to naturally stabilize the weak scale and 
high scale (e.g., GUT scale or Planck scale).  For example, supersymmetry is
nicely compatible with gauge coupling unification and dark matter.
These two good reasons are 
arguably less philosophical than the naturalness reason.  

There has been much discussion recently centered on 
supplanting~\cite{Arkani-Hamed:2004fb} or suspending~\cite{Wells:2003tf} 
our ordinary view of naturalness from consideration in supersymmetry model 
building to allow for very heavy scalar masses.  
A large hierarchy between scalars and fermions is sometimes called
split supersymmetry~\cite{Giudice:2004tc}.   
The phenomenology of split supersymmetry, where all scalars
(even third generation) are significantly heavier than the gaugino
masses, has unique features that put it in contrast with
other approaches to supersymmetry 
(see also~\cite{split pheno 1,split pheno 2}). The ideas of
split supersymmetry may have interesting motivations within
string theory~\cite{split strings}.

In this article, we wish to study the effects of light gauginos
and higgsinos on precision electroweak analysis.  We will demonstrate
below that the best fit to the precision electroweak data, when the
scalar superpartners are decoupled, is light gauginos and higgsinos
near the current direct collider limits.  As a corollary to this
finding,  no combination of gaugino and higgsino masses
above the current direct experimental limits are in conflict with
the precision electroweak data.  This is because as the gauginos
and higgsino get heavier,
the fit approaches the Standard Model fit, which is known to
be compatible with the data as long as the Higgs mass is lighter
than about 200 GeV.  Such a cap on the Higgs boson is guaranteed 
in minimal supersymmetry even if the
superpartner masses decouple to the grand unification 
scale.

We start our analysis by computing the $\chi^2$ fit to the
precision data within the Standard Model using the latest
data and theoretical computations of observables.  We then
review the corrections to the precision observables for
general beyond-the-SM contributions to vector boson
self-energies. We also provide explicit formulas for these 
self-energy functions in the general supersymmetric version of 
the Standard Model~\cite{precision ew groups,precision ew groups 2}.
This is then specialized to 
our case of light gauginos/higgsinos and heavy scalar superpartners.
In the process, we provide some useful
analytic results in the pure gaugino and in the
pure higgsino limits.  The key task within these sections
is to justify the claims made above.  These results and 
additional thoughts are summarized in the conclusions
section.

\section{Standard Model precision electroweak fit}
\setcounter{equation}{0}
\setcounter{footnote}{1}

The Higgs scalar boson has not been found by direct experiment, yet
its effects are present in precision electroweak observables
by virtue of its contributions at one loop to the self
energies of electroweak vector bosons.  In general, new
particles that affect observables only through their induced
oblique corrections, such as the Higgs boson, are best
constrained~\cite{Peskin:2001rw} by the three observables
$\sin^2\theta_{\rm eff}^l$ ($=s^2_{\rm eff}$), $M_W$
and $\Gamma(Z\to l^+l^-)$ ($=\Gamma_l$).

The measurements of these three observables\cite{:2003ih,Renton:2004wd}
are
\begin{eqnarray}
s^2_{\rm eff} &=&0.23147\pm 0.00017~[{\rm Average~ of~ all~ results}],
\label{s2_ave}\\
M_W&=&80.425\pm 0.034~{\rm GeV},\\
\label{mw_ave}
\Gamma_l&=&83.984\pm 0.086~{\rm MeV}.
\label{Gammal_ave}
\end{eqnarray}
It should be noted that the $s^2_{\rm eff}$ determined by the hadronic 
asymmetries
and the one determined by the leptonic asymmetries have a $2.8\, \sigma$ 
discrepancy:
\begin{eqnarray}
s^2_{\rm eff}&=&0.23113\pm 0.00021~[A_{FB}^{0,l},~A_l(P_\tau),~A_l(SLD)],
\label{s2_lep}
\\
s^2_{\rm eff}&=&0.23213\pm 0.00029~[A_{FB}^{0,b},A_{FB}^{0,c},
\langle Q_{FB}\rangle].
\label{s2_had}
\end{eqnarray}
For the Standard Model (SM) analysis, we use the state-of-the-art
computations for $s^2_{\rm eff}$~\cite{Awramik:2004ge},  
$M_W$~\cite{Awramik:2003rn} and
$\Gamma_l$~\cite{Ferroglia:2002rg}.
In the SM, these are expressed in terms of the SM parameters,
$\Delta \alpha_h^{(5)}(M_Z)$, $M_t$, $\alpha_s(M_Z)$ and $M_h$.
For the experimental values of the SM parameters,
we employ the following values: 
\begin{eqnarray}
\Delta \alpha_h^{(5)}(M_Z)&=&0.02769\pm 0.00035~\cite{Jegerlehner:2003rx},
\label{alpha_ave}
\\
M_t&=&178.0\pm 4.3~{\rm GeV}~\cite{Azzi:2004rc},\\
\alpha_s(M_Z)&=&0.1187\pm 0.0020~\cite{RPP}.
\end{eqnarray}
In our analysis, $M_Z$ is fixed to be $91.1875~{\rm GeV}$.

\begin{figure}[t]
\centering
\includegraphics[width=12.0cm,angle=0]{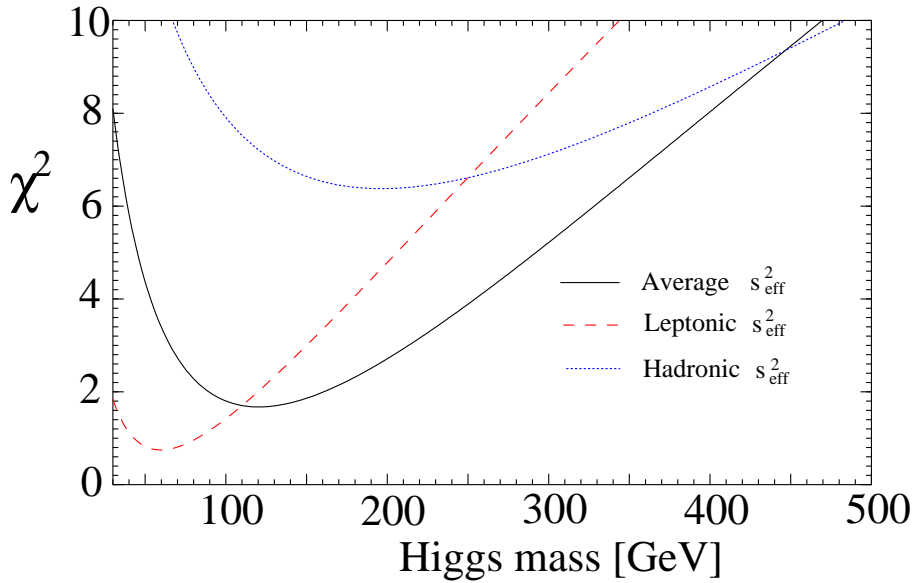}
\caption{Minimum total $\chi^2$ as a function of $M_h$
using different values of $s^2_{\rm eff}$: the average 
$s^2_{\rm eff}=0.23147\pm 0.00017$ (solid line),
the one from leptonic processes only $s^2_{\rm eff}=0.23113\pm 0.00021$ 
(dashed 
line),
and the one from hadronic processes only $s^2_{\rm eff}=0.23213\pm 
0.00029$ 
(dotted line).}
\label{SM_fit}
\end{figure}

By searching for the minimum of $\chi^2$ for the electroweak observables, 
the best-fit Higgs mass can be found. This is carried out by computing
$
\chi^2 (M_h, M_t, \alpha_s, \Delta \alpha_h^{(5)})  = 
\sum_X (X - X_{\rm exp})^2/\sigma_X^2
$
where $X = (M_t,$ $\alpha_s,$ $\Delta \alpha_h^{(5)},$ 
$s^2_{\rm eff},$ $M_W,$ $\Gamma_l)$, with the last three predicted 
in terms of the first four by 
their Standard Model expressions,
and $X_{\rm exp}$, $\sigma_X$ the experimental central values and 
uncertainties.  
In fig.~\ref{SM_fit}, the minimum total $\chi^2$ in the SM is shown as 
a function of $M_h$, with the best-fit values given in Table 
\ref{SM_table}. In all cases, the best fit is achieved for $M_W$ and
$s^2_{\rm eff}$ lower than their experimental central values, and 
$\Gamma_l$ higher than its experimental central value. 

\begin{table}[tbh]
\begin{center}
\begin{tabular}{|c|c|c|c|c|c|c|c|}
\hline 
$\chi^2$ & $M_h$ [GeV] & $M_t$ [GeV] & $\alpha_S(M_Z)$ & 
$\Delta \alpha_{\rm had}^{(5)}(M_Z)$
& $M_W$ [GeV]& $s^2_{\rm eff}$ & $\Gamma_l$ [MeV]\\
\hline \hline
1.67 &  120 & 178.40 & 0.1187 & 0.02775 & 80.391 & 0.23140 & 84.043 \\
 & & ($+0.1\sigma$) & ($0\sigma$) & ($+0.2\sigma$) & ($-1.0\sigma$) & 
($-0.4\sigma$) & ($+0.7\sigma$) \\
\hline
0.75 &  60 & 176.81 & 0.1187 & 0.02772 & 80.418 & 0.23111 & 84.052 \\
 & & ($-0.3\sigma$) & ($0\sigma$) & ($+0.1\sigma$) & ($-0.2\sigma$) & 
($-0.1\sigma$) & ($+0.8\sigma$) \\
\hline
6.38 &  196 & 179.46 & 0.1187 & 0.02784 & 80.366 & 0.23163 & 84.014 \\
 & & ($+0.3\sigma$) & ($0\sigma$) & ($+0.4\sigma$) & ($-1.7\sigma$) & 
($-1.7\sigma$) & ($+0.4\sigma$) \\
\hline
\end{tabular}
\end{center}
\caption{Results of the best fit for the electroweak 
observables in the Standard Model. The first, second, and third rows use 
the averaged,
leptonic, and hadronic values for $s^2_{\rm eff}$, respectively.}
\label{SM_table}
\end{table}

The minimum of the $\chi^2$ is
at $M_h=120$ GeV if the total averaged value of $s^2_{\rm eff}$ is
used in the analysis.  The 95\% confidence level upper bound
on the Higgs mass, which requires $\Delta \chi^2< 1.64$, comes
to $227\gev$. Note that if we used only 
the leptonic data to
compute $s^2_{\rm eff}$, the precision fit would have given
a best fit value for $M_h$ of $60\gev$ and a 95\% CL upper
bound of $132 \gev$.  
If 
we used only 
the hadronic data to
compute $s^2_{\rm eff}$ we obtain a much lower quality fit, with $M_h$ 
of $196\gev$, and a 95\% CL upper bound of $365\gev$.  

Although these differences in the Higgs
mass fit between the
leptonic- and hadronic-determined $s^2_{\rm eff}$ are interesting,
our view at the present is that we should not differentiate the data,
so we take the world averaged $s^2_{\rm eff}$ as the appropriate
observable in our analysis of the Standard Model and its extensions.

\section{Oblique corrections to electroweak observables}
\setcounter{equation}{0}
\setcounter{footnote}{1}

Throughout this paper we will be working in a theoretical framework
where no corrections to electroweak observables are expected through
vertex loops.    Such an assumption is valid for
many theories. No vertex loop corrections
are important if all flavor-charged states that would have contributed
to the vertex corrections are too massive to be relevant.  
In this case, all substantive corrections come from loop corrections
of the vector boson self-energies, which only requires the presence
of light states charged under the symmetries these bosons generate.

In our case, the 
flavor-charged squark and slepton fields are decoupled in split
supersymmetry, suppressing vertex corrections.  However, the gauge-charged
charginos and neutralinos are not decoupled and can contribute substantively
to the electroweak boson self-energies.  That is why we focus on the
oblique corrections.

In this section we give the formalism for general
oblique corrections.  We then apply this formalism to the charginos and
neutralinos of minimal supersymmetry.  We also briefly describe
heavy scalar corrections to the oblique corrections, which will be helpful
in characterizing the small corrections to the chargino/neutralino results
from the heavy scalars sector.

\noindent {\it General oblique corrections}

Our analysis uses the $S,T,U$ parameter expansions
of~\cite{Peskin:1991sw}, augmented by $Y,V,W$ parameters
inspired by~\cite{Maksymyk:1993zm}. The latter take into account the
corrections from nonzero momentum that are important when the new
physics states have mass near $M_Z$. This is crucial in
particular for light charginos and neutralinos in the MSSM, and will
become more important in the future when the top-quark mass, the $W$ boson
mass, and other electroweak observables become known with better accuracy. 

The $S,T,U$ Peskin-Takeuchi parameters are defined as
\begin{eqnarray}
\frac{\alpha S}{4s_W^2 c_W^2}&=&
\frac{\Pi_{ZZ}(M_Z^2)-\Pi_{ZZ}(0)}{M_Z^2}
-\frac{c_{2W}}{c_W s_W} \frac{\Pi_{Z\gamma}(M_Z^2)}{M_Z^2}
-\frac{\Pi_{\gamma \gamma}(M_Z^2)}{M_Z^2},\\
\alpha T&=& \frac{\Pi_{WW}(0)}{M_W^2}-\frac{\Pi_{ZZ}(0)}{M_Z^2},\\
\frac{\alpha U}{4 s_W^2} &=& \frac{\Pi_{WW}(M_W^2)-\Pi_{WW}(0)}{M_W^2}
-c_W^2\frac{\Pi_{ZZ}(M_Z^2)-\Pi_{ZZ}(0)}{M_Z^2}\nonumber \\
&&-2s_W c_W \frac{\Pi_{Z \gamma}(M_Z^2)}{M_Z^2}
-s_W^2 \frac{\Pi_{\gamma \gamma}(M_Z^2)}{M_Z^2}.
\end{eqnarray}
Here we have followed the definitions given, for example, in \cite{RPP}, 
which differ slightly from those given originally in \cite{Peskin:1991sw}.
We use the notation $c_W$ and $s_W$ to refer to the cosine and sine of
the weak mixing angle, and $c_{2W} = c_W^2 - s_W^2$.
All of the self-energy functions $\Pi_{XY}$ are taken to contain only the new 
physics contributions (beyond the Standard Model with a Higgs boson),
and follow the sign convention of e.g. ref.~\cite{RPP}.

In order to completely describe oblique corrections near the $Z$ pole
and at zero momentum, it is necessary to introduce three more parameters, 
as in \cite{Maksymyk:1993zm}. These can be written in combinations $V,W,Y$, 
defined as 
\begin{eqnarray}
\alpha Y &=& \frac{\Pi_{\gamma \gamma}(M_Z^2)}{M_Z^2}
-\hat{\Pi}_{\gamma \gamma}(0),
\\
\alpha V &=& \Pi_{ZZ}^{'}(M_Z^2)-
\left[\frac{\Pi_{ZZ}(M_Z^2)-\Pi_{ZZ}(0)}{M_Z^2}
\right],
\\
\alpha W &=& \Pi_{WW}^{'}(M_Z^2)-
\left[\frac{\Pi_{WW}(M_W^2)-\Pi_{WW}(0)}{M_W^2}
\right],  
\end{eqnarray}
Here $\hat{\Pi}_{\gamma \gamma}(p^2)=\Pi_{\gamma \gamma}(p^2)/p^2$,
and $\Pi'(p^2)=d\Pi/dp^2$.
The parameter $Y$ we use here is a convenient linear combination 
of the parameters originally defined in \cite{Maksymyk:1993zm}. 
Note also that ref.~\cite{Maksymyk:1993zm} used a slightly different 
definition of $S,T,U$ than used here or in other references.

In terms of the parameters defined above, the observables pertinent to our 
discussion 
are expressed as follows: 
\begin{eqnarray}
\frac{M_W^2}{(M_W^2)_{\rm SM}}
&=& 1-\frac{\alpha S}{2 c_{2W}}
+ \frac{c_W^2 \alpha T}{c_{2W}}
+\frac{\alpha U}{4 s^2_W} 
-\frac{s^2_W \alpha Y}{c_{2W}}.\\
\frac{s^2_{\rm eff}}{(s^2_{\rm eff})_{\rm SM}}
&=&1+\frac{\alpha S}{4s^2_W c_{2W} }
-\frac{c_W^2 \alpha T}{c_{2W}} 
+\frac{c_W^2 \alpha Y}{c_{2W}}.\\
\frac{\Gamma_l}{(\Gamma_l)_{\rm SM}}
&=& 1- d_W \alpha S
+ (1+ 4 s_W^2c_W^2 d_W )
\alpha T 
+\alpha V
- 4 s_W^2 c_W^2 d_W \alpha Y,
\end{eqnarray}
where $X_{\rm SM}~(X=M_W,s^2_{\rm eff}~{\rm and}~\Gamma_l)$ are
the SM values, and 
$d_W= (1-4 s^2_W)/[(1-4 s_W^2+8s_W^4) c_{2W}]$.
Note that the quantity $W$ does not contribute at all to these 
particular observables in this parameterization.

\noindent {\it Oblique corrections in low-energy supersymmetry}

The preceding analysis applies to a general theory of new physics in which
vertex corrections are small. In order to employ these parameter
expansions in supersymmetry, we need to compute the contributions to the
vector boson self-energies. These will be given in terms of kinematic
functions $B$, $H$, $G$, $F$, which are defined in the Appendix. They are
implicitly functions of an external momentum invariant $s = p^2$ (in a
signature $+$$-$$-$$-$ metric). In some special cases, it is often
convenient to then expand these kinematic functions in $r = s/M^2$, where
$M$ is the mass of the heavier particle in the loop, to obtain relatively
simple and understandable expressions. The name of a particle stands for
its squared mass when appearing as the argument of a kinematic function.
 
The chargino and neutralino contributions to the electroweak
vector boson self-energies are
\bea
\Pi_{WW} &=& -{g^2 \over 16 \pi^2} \sum_{i=1}^4 \sum_{j=1}^2
\left [
(|O^L_{ij}|^2 + |O^R_{ij}|^2) H(\Ni,\Cj)
+ 4 {\rm Re}[O^L_{ij} O^{R*}_{ij}] M_{\Ni} M_{\Cj} B(\Ni,\Cj)
\right ]
\phantom{xxxx}
\\
\Pi_{ZZ} &=& -{g^2 \over 16 \pi^2 c_W^2} \Bigl \lbrace
\sum_{i,j=1}^4 
\left [
|O^{\prime\prime L}_{ij}|^2 H(\Ni,\Nj)
- 2 {\rm Re}[(O^{\prime\prime L}_{ij})^2] M_{\Ni} M_{\Nj} B(\Ni,\Nj)
\right ]
\nonumber \\ &&
+ 
\sum_{i,j=1}^2
\left [
(|O^{\prime L}_{ij}|^2 + |O^{\prime R}_{ij}|^2 ) H(\Ci,\Cj)
+ 4 {\rm Re}[O^{\prime L}_{ij} O^{\prime R *}_{ij}] M_{\Ci} M_{\Cj} B(\Ci,\Cj) 
\right ]
\Bigr \rbrace
\\
\Pi_{Z\gamma} &=& {g^2 s_W\over 16 \pi^2 c_W}
\sum_{i=1}^2 (O_{ii}^{\prime L} + O_{ii}^{\prime R})
G(\Ci)
\\
\Pi_{\gamma\gamma} &=& -{g^2 s_W^2 \over 8 \pi^2}
\sum_{i=1}^2
G(\Ci)
\eea
The notation for the chargino and neutralino couplings is the
same as in \cite{conventions,Martin:2002iu}, and can be described as 
follows.
In the $(\stilde B, \stilde W^0, \stilde H_d^0, \stilde
H_u^0)$ basis, the neutralino mass matrix is
\beq
M_{\tilde \chi^0} = \pmatrix{M_1 & 0 & -\gp \vd/\sqrt{2} & \gp \vu/\sqrt{2} \cr
                         0 & M_2 & g \vd/\sqrt{2} & - g\vu /\sqrt{2} \cr
                        -\gp \vd/\sqrt{2} & g \vd/\sqrt{2} & 0 & -\mu \cr
                        \gp \vu/\sqrt{2} & - g\vu /\sqrt{2} & -\mu & 0},
\label{neutmass}
\eeq
where $v_{u,d}=\langle H^0_{u,d}\rangle$ and $v_u/v_d=\tan\beta$, such that
$v_u^2+v_d^2\simeq (174 \gev)^2$.
The unitary matrix $N$ diagonalizes $M_{\tilde\chi^0}$:
\beq
N^* M_{\tilde \chi^0} N^{-1} &=& 
{\rm diag}( M_{\tilde \chi^0_1}, M_{\tilde\chi^0_2}, M_{\tilde \chi^0_3}, 
M_{\tilde \chi^0_4}),
\label{diagonalizemN}
\eeq
where the mass eigenvalues $M_{\tilde \chi^0_i}$ are all real and 
positive (see~\cite{Martin:2002iu} for technique).
In the $(\stilde W^\pm,\stilde H^\pm)$ basis, the chargino mass matrix is
\beq
M_{\tilde \chi^+} = \pmatrix{M_2 & g\vu \cr
                        g \vd & \mu}.
\label{charmass}
\eeq
The unitary matrices $U$ and $V$ diagonalize the above matrix according to
\beq
U^* M_{\tilde \chi^+} V^\dagger &=& 
\pmatrix{M_{\tilde \chi^+_1} & 0 \cr
           0 & M_{\tilde \chi^+_2}}
\eeq
where again $M_{\tilde \chi^+_i}$ are real and positive. 
One finds $U$ and $V$
by solving
\beq
VM_{\tilde \chi^+}^\dagger M_{\tilde \chi^+}V^{-1} =
U M_{\tilde \chi^+}^* M_{\tilde \chi^+}^T U^{-1} = 
\pmatrix{M_{\tilde \chi^+_1}^2 & 0 \cr
           0 & M_{\tilde \chi^+_2}^2}.
\eeq
The $O_{ij}$-couplings are
\beq
&&O^{L}_{ij} = N_{i2} V_{j1}^* - N_{i4} V_{j2}^*/\sqrt{2},
\qquad\qquad\>\>\>\>\>\>
O^{R}_{ij} = N^*_{i2} U_{j1} + N^*_{i3} U_{j2}/\sqrt{2}, \nonumber
\\
&&O^{'L}_{ij} = -V_{i1} V_{j1}^* - {1\over 2} V_{i2} V_{j2}^* +
s_W^2\delta_{ij},
\qquad\>\>\>
O^{'R}_{ij} = -U_{i1}^* U_{j1} - {1\over 2} U_{i2}^* U_{j2} +
s_W^2\delta_{ij}, \nonumber
\phantom{xxx}
\\
&&O^{''L}_{ij} = (-N_{i3} N_{j3}^* + N_{i4} N_{j4}^*)/2
.  
\eeq
In eqs.~(\ref{neutmass}) and (\ref{charmass}), 
we have assumed that the gaugino couplings
to Higgs-higgsino pairs are given by the tree-level supersymmetric relation.
We will discuss the merits of this assumption in section~\ref{sec:inofits}.

For completeness, we also compute oblique corrections due to the
sfermions and heavy Higgs bosons.  As we stated in the introduction,
we are assuming that the sfermions and heavy Higgs bosons are decoupled
and have no substantive effect on the fits if their masses are 
above a TeV.  The equations below are used to justify that statement.

First, for the sfermions,
we assume, as is consistently
suggested by experimental constraints and theoretical 
prejudice,
that the first two families have negligible sfermion mixing. 
For the third family sfermions $\tilde t_i$, $\tilde b_i$, and $\tilde
\tau_i$ with $i=1,2$, the mixing (including possible CP violating phases)
is described by 
\beq
\pmatrix{\tilde f_L \cr \tilde f_R} = 
\pmatrix{ c_{\tilde f} & -s_{\tilde f}^* \cr s_{\tilde f} &c_{\tilde f}^*} 
\pmatrix{\tilde f_1 \cr \tilde f_2}  
\eeq
where $|c_{\tilde f}|^2 + |s_{\tilde f}|^2 = 1$. When there is no CP
violation, $c_{\tilde f}$ and $s_{\tilde f}$ are real and are the sine and
cosine of a sfermion mixing angle. 
Then the sfermion contributions to the self energies of the vector bosons 
are 
\beq 
\Pi_{WW} &=& {g^2 \over 32 \pi^2} \Bigl [
3 F(\tilde d_L, \tilde u_L) + 3  F(\tilde s_L, \tilde c_L) 
+ 3 \sum_{i,j = 1}^2 | g_{W \tilde b_i \tilde t_j^*} |^2 
F(\tilde b_i, \tilde t_j)
\nonumber \\ &&
+ F(\tilde e_L, \tilde \nu_e) + F(\tilde \mu_L, \tilde \nu_\mu) 
+ \sum_{i=1}^2 | g_{W \tilde \tau_i \tilde \nu_\tau^*} |^2 
F(\tilde \tau_i,\tilde \nu_\tau)
\Bigr ]
\\
\Pi_{ZZ} &=& {g^2 \over 16 \pi^2 c_W^2} \sum_f N_f 
\sum_{i,j} |g_{Z\tilde f_i \tilde f_j^*}|^2
                  F(\tilde f_i, \tilde f_j)
\label{PIZZsfermions} 
\\
\Pi_{Z\gamma} &=& {g^2 s_W \over 16\pi^2 c_W} \sum_{\tilde f_i} 
N_f Q_f g_{Z\tilde f_i\tilde f_i^*} F(\tilde f_i, \tilde f_i)
\label{PIZgsfermions}
\\
\Pi_{\gamma\gamma} &=& {g^2 s_W^2 \over 16\pi^2 } \sum_{\tilde f_i} 
N_f Q_f^2 F(\tilde f_i, \tilde f_i)
\label{PIggsfermions}
\eeq
Here, $N_f=3,1$ and $Q_f = +2/3, -1/3,
-1,0$ in the obvious way. In eq.~(\ref{PIZZsfermions}) the sum on $f$ is
over the 12 symbols $(d,s,b,u,c,t,e,\mu,\tau,\nu_e,\nu_\mu,\nu_\tau)$, and
$i,j$ run over 1,2, except for the sneutrinos. In
eqs.~(\ref{PIZgsfermions}) and (\ref{PIggsfermions}), the sums are over
the 18 charged sfermion mass eigenstates. The $Z$ couplings for the first
two families and the tau sneutrino are 
\beq
&&g_{Z\tilde d_L \tilde d_L^*} = -{1/2} + s_W^2/3,
\qquad
g_{Z\tilde u_L \tilde u_L^*} = {1/2} - 2s_W^2/3,
\qquad
g_{Z\tilde e_L \tilde e_L^*} = -{1/2} + s_W^2, \nonumber
\phantom{xxxxx}
\\
&&g_{Z\tilde \nu\tilde \nu^*} = {1/2},
\qquad
g_{Z\tilde u_R \tilde u_R^*} = -2 s_W^2/3,
\qquad
g_{Z\tilde d_R \tilde d_R^*} =  s_W^2/3,
\qquad
g_{Z\tilde e_R \tilde e_R^*} =  s_W^2, 
\eeq
and for the third-family sfermions other than the tau sneutrino are
\beq
&&g_{Z\tilde f_1 \tilde f_1^*} = 
|c_{\tilde f}|^2 g_{Z\tilde f_L \tilde f_L^*} 
+ |s_{\tilde f}|^2 g_{Z\tilde f_R \tilde f_R^*},
\qquad\qquad
g_{Z\tilde f_2 \tilde f_2^*} = 
|s_{\tilde f}|^2 g_{Z\tilde f_L \tilde f_L^*} 
+ |c_{\tilde f}|^2 g_{Z\tilde f_R \tilde f_R^*}, \nonumber
\phantom{xxxxx}
\\
&& g_{Z\tilde f_1 \tilde f_2^*}  = (g_{Z\tilde f_2 \tilde f_1^*})^*
= s_{\tilde f} c_{\tilde f} (g_{Z\tilde f_R \tilde f_R^*} 
- g_{Z\tilde f_L \tilde f_L^*}). 
\eeq
The $W$ couplings for the third-family sfermions are
\beq
&&g_{W\tilde b_1\tilde t_1^*} = c_{\tilde b} c_{\tilde t}^*,
\qquad
g_{W\tilde b_2\tilde t_2^*} = s_{\tilde b}^* s_{\tilde t},
\qquad
g_{W\tilde b_1\tilde t_2^*} = -c_{\tilde b} s_{\tilde t},
\qquad
g_{W\tilde b_2\tilde t_1^*} = -s_{\tilde b}^* c_{\tilde t}^*, \nonumber
\phantom{xxx} 
\\
&&g_{W \tilde \tau_1\tilde \nu_\tau} = c_{\tilde \tau},
\qquad
g_{W\tilde \tau_2\tilde \nu_\tau} = -s_{\tilde \tau}^*.
\eeq

Finally, we consider the contributions of the Higgs scalar
bosons, $h^0$, $H^0$, $A^0$, and $H^\pm$.
We assume that the Standard Model result already includes
contributions from the lightest Higgs scalar. Therefore, to compensate, in
the following we subtract a contribution from $h^0$ with Standard Model
couplings, in other words $\sin^2(\beta-\alpha) \rightarrow 1$. This just
converts each term involving $h^0$ and $W,Z$ with coefficient
$\sin^2(\beta-\alpha)$ into one with coefficient $-\cos^2(\beta-\alpha)$.
The results below are therefore the difference between the MSSM and the
Standard Model with Higgs mass $m_{h}$: 
\beq
\Pi_{WW} &=& {g^2 \over 64\pi^2} \Bigl \lbrace 
F(A^0, H^+) + \sin^2(\beta-\alpha) F(H^0, H^+) 
+\cos^2(\beta-\alpha) \bigl [ F(h^0,H^+)  
\nonumber \\ &&
+ F(H^0,W) - 4 M_W^2 B(H^0,W) - F(h^0,W) + 4 M_W^2 B(h^0,W)]
\Bigr \rbrace
\\
\Pi_{ZZ} &=& {g^2 \over 64 \pi^2 c_W^2}
\Bigl \lbrace
c_{2W}^2 F(H^+,H^+) + \sin^2(\beta-\alpha) F(A^0,H^0)
+ \cos^2 (\beta-\alpha) \bigl [
F(h^0,A^0) 
\phantom{xxx}
\nonumber \\ &&
+ F(H^0,Z) - 4 M_Z^2 B(H^0,Z) - F(h^0,Z) + 4 M_Z^2 B(h^0,Z) \bigr ]
\Bigr \rbrace
\\
\Pi_{Z\gamma} &=& {g^2 s_W c_{2W} \over 32 \pi^2 c_W}  
F(H^+,H^+)
\\ 
\Pi_{\gamma\gamma} &=& {g^2 s_W^2 \over 16 \pi^2} F(H^+,H^+) .
\eeq

\section{Precision fits with light supersymmetric fermions\label{sec:inofits}}
\setcounter{equation}{0}
\setcounter{footnote}{1}

We are now in position to compute the effects of light supersymmetric
particles and Higgs scalars on precision electroweak observables.  In this
section we consider the effects of light supersymmetric
fermions (charginos and 
neutralinos). 
We do this in the light gaugino limit first, then the
light higgsino limit.  These two limits admit nice analytic results, and
are interesting to study in their own right.  We then consider results for
a general mixed higgsino and gaugino scenario.

\noindent {\it Light gauginos limit}

In this section we assume that the $\mu$ term is very heavy along with
the squarks and sleptons.  Thus, we assume all states in the theory
decouple except a light SM-like Higgs boson and light gauginos.
We then have pure electroweak bino and winos at the low-energy scale.
The gluino can also be light, of course, but it has no effect
on the electroweak observables.
Thus,
all results depend on the unknown $h$ mass and the unknown 
wino mass.  
Since the pure bino does not couple to $Z$, $W$, and $\gamma$, it does
not contribute to oblique corrections, and so
its value with respect to the wino mass is irrelevant to this
analysis.

The values of the couplings for the 
light gaugino limiting case are 
\bea
O^L_{21} = O^R_{21} = 
-O^{\prime L}_{11}/c_W^2 =  -O^{\prime R}_{11}/c_W^2 = 1,
\eea
and all other couplings are either zero or irrelevant.
The expressions for the self-energies in this case collapse into 
a rather convenient form:
\beq
\Pi_{WW} = \frac{\Pi_{ZZ}}{c_W^2} = 
\frac{\Pi_{Z\gamma}}{s_W c_W} = 
\frac{\Pi_{\gamma\gamma}}{s_W^2} = 
- {g^2 \over 8 \pi^2} G(\tilde W).
\label{wino pis}
\eeq
The function $G(x)$ is defined in the appendix. The $p^2$ argument of
$G(x)$, and of other loop functions that we will define later, 
is not explicitly
written for simplicity of notation.  

We note immediately
that $S=T=0$ for this case. Thus, an $S-T$ parameter analysis cannot
capture the effects of winos on precision electroweak observables.
The non-zero contributions to the precision
electroweak observables come from the $U$, $Y$ and $V$ parameters.
These parameters can be computed straightforwardly given their definitions
and eq.~(\ref{wino pis}).  We use these to obtain
a convenient expansion of the observables in powers of 
$\rwino=M_Z^2/M^2_{2}$: 
\beq
\Delta M_W({\rm GeV}) &=& 
          0.00954\, \rwino + 0.00157\, \rwino^2 + 0.00030\, \rwino^3, \\
\Delta s^2_{\rm eff} &=& -0.0000549\, \rwino - 0.0000059\,\rwino^2 - 
         0.0000009\,\rwino^3, \\
\Delta \Gamma_l({\rm MeV}) &=& 
          -0.0435\,\rwino - 0.0096\,\rwino^2 - 0.0022\,\rwino^3 .
\eeq
We have found numerically that this expansion is quite accurate even when
$\rwino$ is near one, provided that $|\mu|$ is large. 

Note that in the above equations
$\Delta M_W >0$, $\Delta s^2_{\rm eff}<0$ and $\Delta \Gamma_l<0$
for the wino corrections.
Comparing the SM predictions in Table~\ref{SM_table} with the 
experimental data in eqs.~(\ref{s2_ave})-(\ref{Gammal_ave}), 
the light winos improve
$M_W$ and $\Gamma_l$ predictions.
To see how light winos can affect these observables, 
we consider ``0.5-$\sigma$
sensitive wino mass'' which changes the SM predictions of observables
by 0.5-$\sigma$. The current experimental uncertainties for observables
are $\delta M_W=0.034$ GeV, $\delta s^2_{\rm eff}=0.00017$ and
$\delta \Gamma_l =0.086$ MeV at 
1-$\sigma$ level, and hence we can calculate
``0.5-$\sigma$ sensitive wino mass'' using the above expansions. They are
$M_2=77$ GeV from $\Delta M_W=\delta M_W/2$, 
$M_2=79$ GeV from $\Delta s^2_{\rm eff}=\delta s^2_{\rm eff}/2$ and
$M_2=101$ GeV from $\Delta \Gamma_l=\delta \Gamma_l/2$. 
Therefore winos with $M_2 \sim 100$ GeV
can affect the electroweak fit. Note that $\Gamma_l$ is the most sensitive to
the light winos and it is improved. 

\begin{figure}[t]
\centering
\includegraphics[width=13.0cm,angle=0]{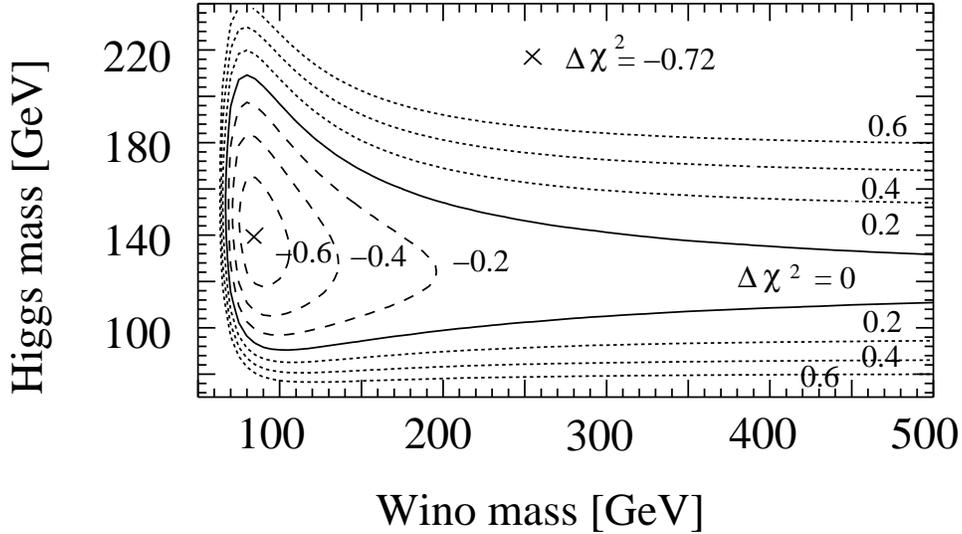}
\caption{Contours of $\Delta\chi^2=\chi^2-\chi^2_{SM,min}$ 
as a function of Higgs mass 
$M_h$ and the
wino mass $M_2$. The region inside (outside) of the solid line produces
a better (worse) fit than the best fit point of the Standard Model, which 
corresponds to the
limit of superpartner masses decoupling to infinity 
and $M_h = 120$ GeV. The best fit point, 
not taking
into account direct searches for the charged wino, is indicated by
the point marked $\times$ at $M_h = 141$ GeV, $M_2 = 86$ GeV.
This figure was made using $\alpha_s(M_Z)=0.1187$.}
\label{wino_fit}
\end{figure}

In fig.~\ref{wino_fit}, we show the total 
$\Delta\chi^2=\chi^2-\chi^2_{SM,min}$ for the electroweak
fit as a function of Higgs mass $M_h$ and wino mass $M_2$.
As one can see, the $\chi^2$ improves as $M_2$ gets smaller 
(up to about 90 GeV) in the range of
115 GeV $<M_h<$ 170 GeV.
The minimum of the $\chi^2$ is about $0.95$ at $M_h\simeq 140$ GeV
and $M_2\simeq 85$ GeV. Note that the minimum of the $\chi^2$
in the SM is about $\chi^2_{SM,min}=1.7$ 
at $M_h=120$ GeV (see fig.~\ref{SM_fit}
and Table~\ref{SM_table}).
Note also that the current wino mass limit
is about $90$ GeV when all squarks and higgsinos are 
heavy~\cite{Heister:2002mn}.

\noindent {\it Light higgsinos limit}

Next, we consider a different limit: all gaugino masses are
large, but $\mu$ is small. In this case, low-energy charginos and
neutralinos are pure higgsinos.  All precision electroweak results
are functions of the light SM-like Higgs mass and the higgsino
mass $\mu$.

The light higgsino case corresponds to 
$O^L_{11} = O^R_{11} = -1/2$, and $O^L_{21} = O^R_{21} = i/2$,
$O^{\prime L}_{11} = O^{\prime R}_{11} = s_W^2 - 1/2$,
$O^{\prime\prime L}_{11} = O^{\prime\prime L}_{22} = 0$, and
$O^{\prime\prime L}_{12} = -O^{\prime\prime L}_{21} = i/2$, 
and all others are irrelevant.
The vector boson self energies reduce to
\beq
\Pi_{WW} = {c_W^2 \over c_W^4 + s_W^4} \Pi_{ZZ} =
{c_W \over s_W (c_W^2 - s_W^2)} \Pi_{Z\gamma} = 
{1\over 2 s_W^2} \Pi_{\gamma\gamma} =  
- {g^2 \over 16\pi^2}G(\tilde H)
\eeq
Similar to the pure gaugino limit, $S=T=0$ in this limit,
and $S-T$ analysis alone cannot capture
the effects of light higgsinos on precision electroweak observables.

We can also expand the pure higgsino limit analytically as a power
series in $\rmu=m^2_Z/\mu^2$:
\beq
\Delta M_W({\rm GeV}) &=& 0.00620\,\rmu + 0.00094\,\rmu^2 + 
0.00017\,\rmu^3,\\
\Delta s^2_{\rm eff} &=& -0.0000549\,\rmu - 0.0000059\,\rmu^2 
- 0.0000009\,\rmu^3,\\
\Delta \Gamma_l({\rm MeV}) &=& -0.0225\,\rmu - 0.0051\,\rmu^2 
- 0.0012\,\rmu^3 .
\eeq

\begin{figure}[t]
\centering
\includegraphics[width=13.0cm,angle=0]{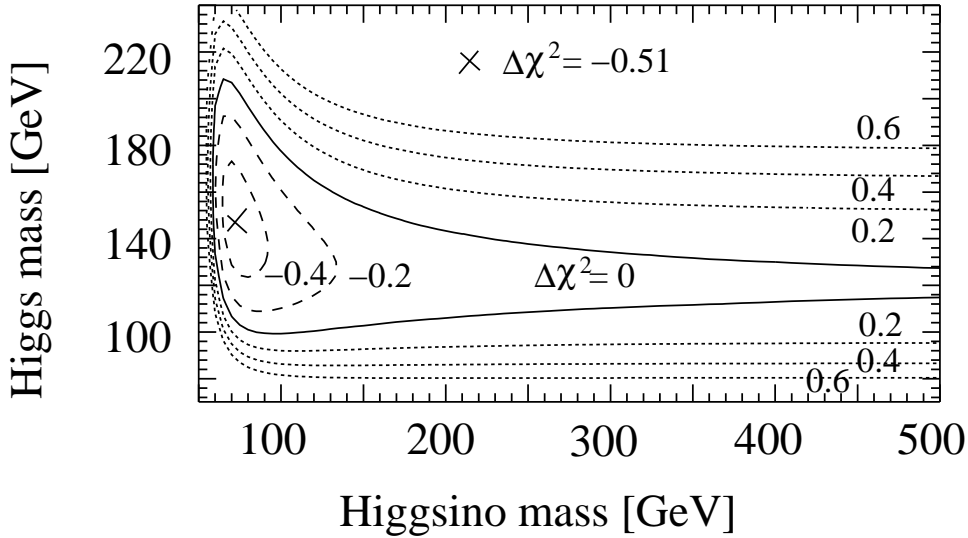}
\caption{Contours of $\Delta\chi^2=\chi^2-\chi^2_{SM,min}$ 
as a function of Higgs mass 
$M_h$ and the
higgsino mass. The region inside (outside) of the solid line produces
a better (worse) fit than the best fit point of the Standard Model, which 
corresponds to the
limit of all superpartners decoupling to infinite mass and
$M_h = 120$ GeV. The best fit point, 
not taking
into account direct searches for the charged higgsino, is indicated by
the point marked $\times$ at Higgs mass of $147\gev$ and higgsino mass
of $73\gev$. This figure was made using $\alpha_s(M_Z)=0.1187$.}
\label{higgsino_fit}
\end{figure}

The higgsinos also improve $M_W$ and $\Gamma_l$ compared to the SM 
predictions. Again, we can calculate ``0.5-$\sigma$ sensitive 
higgsino mass'': $\mu=65$ GeV from $\Delta M_W=\delta M_W/2$, 
$\mu=79$ GeV from $\Delta s^2_{\rm eff}=\delta s^2_{\rm eff}/2$
and $\mu=78$ GeV from $\Delta \Gamma_l=\delta \Gamma_l/2$.  
(Recall from above that $\delta {\cal O}_i$ is the $1\sigma$
experimental error for observable ${\cal O}_i$.)
The total $\Delta\chi^2$ 
is shown in fig.~\ref{higgsino_fit} 
as a function of $\mu$ and Higgs mass $M_h$. 
The effect is smaller than what we found in the pure wino case, but
the light higgsinos (with $\mu>80$ GeV)
also improve the total $\chi^2$ compared to the SM fit.

\noindent {\it Mixed gauginos and higgsinos}

Here we consider the more general case for light charginos and
neutralinos. When the $\mu$-term is near in mass to the wino
mass term $M_2$, both the higgsino and wino sectors contribute
substantially to the oblique corrections. In this case the
general mixing matrix angles for the charginos and neutralinos
vary over large ranges.  Unfortunately,
there are no simple analytic equations that capture all the
effects succinctly. All results for the mixed case will be
numerical results taking into account proper diagonalizations
of the mass matrices.

Unlike the pure gaugino and the pure higgsino cases, the mixed-case
results strongly depend on $\tan\beta$.  This is easy to understand since
the gaugino/higgsino mass mixing insertions are $\tan\beta$ dependent
and not small.  Thus, $\tan\beta$ is a crucial parameter to keep
track of.  
Another parameter that we should keep track of is the ratio of the higgsino
parameter to the wino parameter, $\mu/M_2$.  We will assume that this 
mixing parameter can be anything along the real line (no complex, CP violating
phases). 

\begin{figure}
\centering
\includegraphics[width=11.0cm,angle=0]{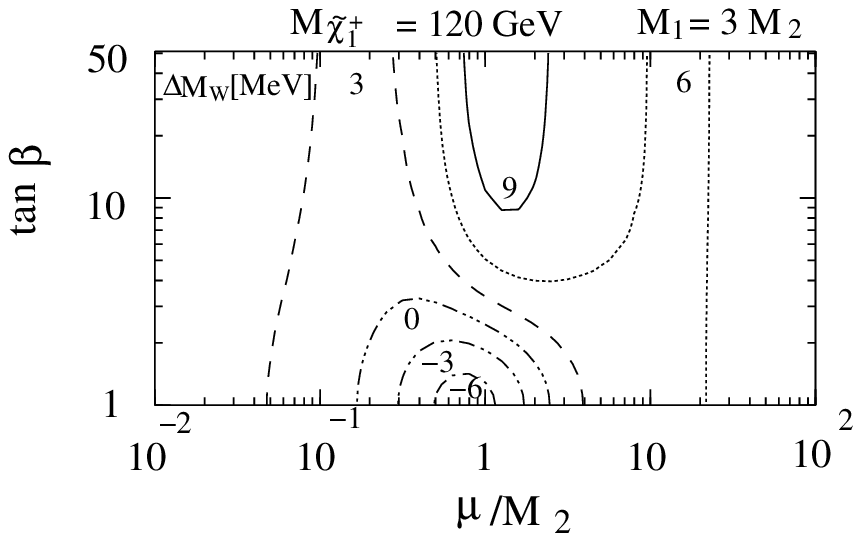}
\includegraphics[width=11.0cm,angle=0]{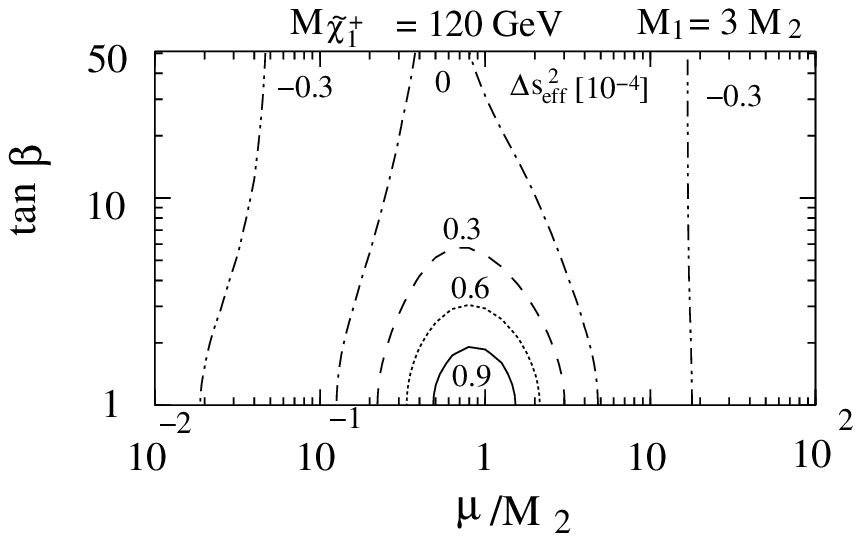}
\includegraphics[width=11.0cm,angle=0]{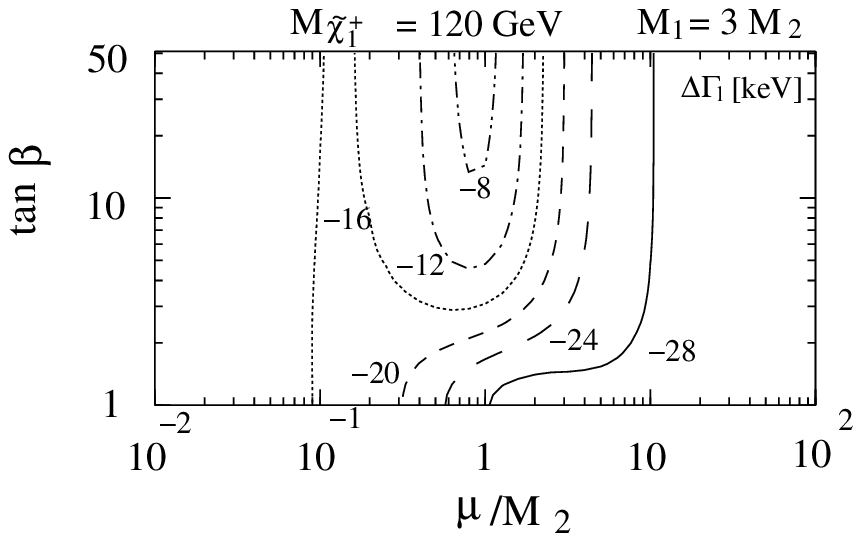}
\caption{Contours of shifts in electroweak observables 
in the $\tan\beta$ vs. $\mu/M_2$ plane, with
the lightest chargino mass eigenstate fixed at $120\gev$
and $M_1=3M_2$.}
\label{d_obs_AMSB}
\end{figure}

We demonstrate the effect of mixing by plotting contours of shifts
in observables in the $\tan\beta$ vs. $\mu/M_2$ plane.  We do this
in fig.~\ref{d_obs_AMSB} keeping the lightest chargino mass eigenvalue
fixed at $M_{\tilde \chi^+_1}=120\gev$. We also assume that $M_1\simeq 3M_2$
according to anomaly mediated supersymmetry~\cite{AMSB}, 
although we have checked that
the results depend very mildly on this assumption.  Choosing $M_1\simeq M_2$
or $M_1\simeq M_2/2$,
for example, generates very similar figures.

There are several things to notice in fig.~\ref{d_obs_AMSB}. First, as
$\mu/M_2\gg 1$ the precision electroweak analysis asymptotes to that of light
wino superpartners.  As $\mu/M_2\ll 1$ the precision 
electroweak analysis asymptotes
to that of light higgsinos.  In both cases, the $\tan\beta$ dependence
disappears and the variation of the corrections to the observables
disappears when there is a factor of 10 or higher in the hierarchy of
$\mu$ and $M_2$. 
One finds in fig.~\ref{d_obs_AMSB} a strong $\tan\beta$ dependence 
when $\mu/M_2\sim 1$, which can induce a correction of
either sign for $\Delta M_W$,
depending on the value of $\tan\beta$, and
only positive (negative) corrections to $s^2_{\rm eff}$ ($\Gamma_l$).
The variability in the corrections is large in that region.

\begin{figure}
\centering
\includegraphics[width=11.0cm,angle=0]{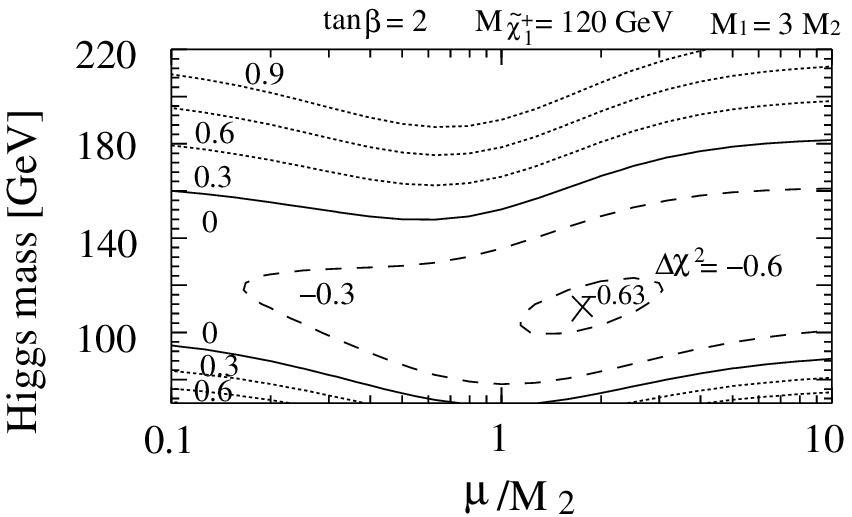}
\includegraphics[width=11.0cm,angle=0]{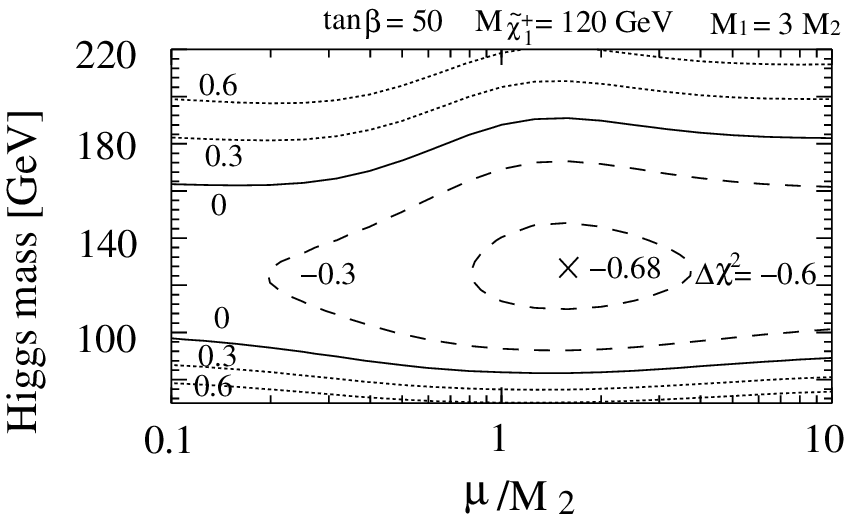}
\caption{$\Delta\chi^2$ contours with $\tan\beta=2,50$ and
$M_{\tilde\chi^+_1}=120\gev$ fixed.  We have also chosen  $M_1=3M_2$,
although the contours depend very mildly on this assumption.  The results
illustrate the general finding that a mixed scenario of light
gauginos and higgsinos lead to a better fit to the precision
electroweak data. This figure was made using $\alpha_s(M_Z)=0.1187$.} 
\label{d_chi2_AMSB}
\end{figure}

Having established that the observables change significantly when
$\mu \simeq M_2$, we now wish to determine the effect these variations
have on $\Delta\chi^2$.  To this end we have plotted 
in fig.~\ref{d_chi2_AMSB} contours of
$\Delta \chi^2$ in the plane of $M_{h}$ versus $\mu/M_2$.
We have fixed $M_{\tilde \chi^+_1}=120\gev$, and $\tan\beta=2$
in the top graph and $\tan\beta=50$ in the bottom graph.

On the far left of the figure we have the result of the pure higgsino
case, where the $\Delta \chi^2$ changes only when the Higgs boson mass
changes.  The contour lines become parallel to the $x$-axis.  On the far
right of figure we find the result asymptoting toward the pure gaugino case.
Again, $\Delta \chi^2$ changes only due to the Higgs boson mass, and the
lines again level horizontally out there.
In the center of the contours of fig.~\ref{d_chi2_AMSB} the mixing
angles are varying significantly.  The variation is causing the changes
in the observables that we witnessed in fig.~\ref{d_obs_AMSB},
and subsequently affects $\Delta\chi^2$.  As we see, the mixing effect
reduces $\Delta\chi^2$.  

Reduction of $\Delta\chi^2$ in the region of parameter
space where gauginos and higgsinos are light and heavily mixed
is a general result in split supersymmetry.  We can understand this
result from the graphs.  Let us draw our attention to a segment of 
the graphs at $\mu/M_2\sim 1$.  At high $\tan\beta$ the corrections to
$s^2_{\rm eff}$ and $\Gamma_l$ are becoming small, whereas the correction
to $M_W$ is increasing.  The increasing contribution to $M_W$ makes
the theory prediction come closer to the experimental prediction, thus
reducing the $\Delta\chi^2$.  At low $\tan\beta$ the contribution to
$M_W$ is negative. Although this goes in the wrong direction, the magnitude
is somewhat smaller than in the large $\tan\beta$ case, and more importantly,
the contributions to $s^2_{\rm eff}$ are increasingly positive, which
goes in the right direction, and the contributions to $\Gamma_l$
are increasingly negative, which also goes in the right direction.
Overall, the $\chi^2$ improves.  This result is true for all $\tan\beta$
although we have only shown it graphically for $\tan\beta=2$ and 
$\tan\beta=50$.  The improved $\Delta \chi^2$
results also hold similarly for $\mu/M_2\simeq -1$.
Therefore, mixed higgsinos and gauginos near the direct
experimental limit is the split supersymmetry spectrum most
compatible with the precision electroweak data.

Let us say a few words about dark 
matter~\cite{Wells:2003tf,split pheno 1,split pheno 2} in relation to
our precision electroweak analysis.  If we assume that R-parity
is conserved, the lightest supersymmetric partner (LSP) will be stable.
In the case of a pure
wino LSP, the dark matter thermal relic
abundance is negligible unless the mass is about $2.3\tev$.  Pure
higgsino LSP has negligible relic abundance unless its mass
is about $1.2\tev$.  Higher masses mean overclosure, i.e., cosmological
problems. However,
these conclusions are applicable for thermal relic abundance
calculations.  Non-thermal sources, such as gravitino or moduli
decay in the early universe can transform what looked to be a negligibly
abundant LSP into a good dark matter candidate.  Thus, the light
winos, light higgsinos and light mixed states are probably not good
thermal dark matter candidates, but could be good dark matter candidates
when all non-thermal sources are taken into account.  If $M_1<M_2$ and
the LSP has significant bino fraction, one expects either
the LSP to annihilate efficiently through a Higgs boson pole
or $\mu$ should be somewhat near 
$M_1$ to mix with the bino for acceptable dark matter (see Pierce
in~\cite{split pheno 1}).  
This is good for dark matter and
good for precision electroweak fits.

\noindent {\it Indirect effects from heavy scalars}

In the above analysis we have assumed that the neutralino
and chargino matrices used to obtain the mass eigenvalues and
mixing angles are valid.  However, the $g$ and $g'$
that are in those matrices are gauge couplings because of supersymmetry
invariance.  In broken supersymmetry those couplings deviate from
the gauge couplings, which has been emphasized within the context
of split supersymmetry~\cite{Arkani-Hamed:2004fb,Giudice:2004tc}.

Unfortunately, these ``gauge-ino couplings'' cannot be directly
measured by experiment since the scalar masses are likely to 
not be accessible. However, they can have a subtle effect on the
precision electroweak observables.  To demonstrate, we introduce
the following couplings
\beq
&a_u = \tilde g_u/(g\sin\beta),~~\qquad &a_d=\tilde g_d/(g\cos\beta)  \\ 
&a_u'=\tilde g'_u/(g'\sin\beta),~~\qquad &a'_d=\tilde g'_d/(g'\cos\beta) 
\eeq
where $\tilde g_{u,d}$ and $\tilde g'_{u,d}$ are defined 
in~\cite{Giudice:2004tc}. 
The $a$-variables are defined such that the usual values taken in the
MSSM are $a_u=a_d=a'_u=a'_d=1$.  

The neutralino and chargino mass matrices at the weak
scale in this parameterization are
\bea
M_{\tilde \chi^0}=\left( \begin{array}{cccc}
M_1 & 0 & -a'_dg'v_d/\sqrt{2} & a'_ug'v_u/\sqrt{2} \\
0 & M_2 & a_dgv_d/\sqrt{2} & -a_ugv_u/\sqrt{2} \\
-a'_dg'v_d/\sqrt{2} & a_dgv_d/\sqrt{2} & 0 & -\mu \\
a'_ug'v_u/\sqrt{2} & -a_ugv_u/\sqrt{2} & -\mu & 0
\end{array}\right),
\eea
\bea
M_{\tilde \chi^+} =\left( \begin{array}{cc}
M_2 & a_u g v_u \\
a_d g v_d & \mu 
\end{array}\right)
\eea

We have computed the numerical values of the $a$-variables under
different assumptions for $\tan\beta$ and the scale of the 
scalar superpartners. (We have reproduced fig. 5 of~\cite{Giudice:2004tc}
and agree with their results.)  We then compute the corrections to
the precision electroweak observables for various values of the
scalar sector mass, which for simplicity we assume is a common
scale $M_s$. For PeV-scale sfermion masses none of the $a$-variables
deviate from 1 by more than 10\% for any value of $\tan\beta$.  
If the scalar masses are near the GUT scale, the effects can be
more sizable and deviations from $a_i=1$ can approach $30\%$
at very high $\tan\beta\sim 50$, but are less significant for
lower $\tan\beta$.

We note that the effects of various $M_s$ values can cause the
magnitude of the oblique correction to change by as much as
$30\%$ at special points such as when
$M_2\simeq |\mu| \simeq 100\gev$ and $M_s\simeq 10^{16}\gev$.  
This is especially
true for $s^2_{\rm eff}$, which is very sensitive to the chargino
and neutralino mixing angles. However, when $\mu$ deviates from $M_2$
one finds the effects on precision electroweak corrections
to be much smaller.  In all cases, the corrections to the oblique
corrections are not discernible by current experiment.
This is why we ignored these $a'_{u,d}$ and $a_{u,d}$
parameters for much of our analysis.  In the future, these small deviations
might be discernible at dedicated next-generation $Z$ 
factories~\cite{GigaZ}.


\section{Conclusion}
\setcounter{equation}{0}
\setcounter{footnote}{1}

We have found above that 
the precision electroweak corrections from light charginos and
neutralinos generally improve the overall $\chi^2$ fit to the
data.  This is true in the pure gaugino limit and in the pure
higgsino limit. We emphasize that in both of these cases we have
merely added one new parameter to the theory and the fit gets
better.  It did not have to go this direction.  In the
mixed higgsino and gaugino case, there are more free parameters
introduced, and it turns out that the fit to the data gets even
better.  The case of 
higgsinos and gauginos both near the direct experimental limit
($\sim 100\gev$) is the split supersymmetry spectrum most compatible with the
data.

These conclusions are made within the full supersymmetric 
framework, where we have assumed the scalar superpartner
masses are too heavy to have any noticeable effect on
the precision electroweak observables.  Even for
scalar superpartners decoupled up to the grand unification
scale, the light Higgs mass is still less than about
$170\gev$~\cite{Giudice:2004tc}.  Our best fit results are all compatible with
this low range of Higgs mass, and the global fits for the
various cases we discussed have global minima with light
Higgs boson mass.
The global fit to the data approaches that of the SM fit
when the gaugino and higgsino masses are dialed to larger
values.  In that case, the global minimum of the $\chi^2$
fit remains in the low Higgs mass region, but the $\chi^2$
value at that SM minimum is increased somewhat compared to the
light gauginos/higgsino case.

Throughout we have assumed that the squarks, sleptons and
heavy Higgs bosons have no effect on precision electroweak
analysis.  Our computations demonstrate that we expect less
than a $10^{-4}$ effect on all relative corrections
of the observables,
$\Delta {\cal O}_i/{\cal O}_i$, if the scalar masses are
above $1\tev$. In split supersymmetry, we expect the scalar
masses to be significantly beyond the TeV scale, justifying our
neglect of the scalar masses. As an aside, we have 
found that lowering the sfermion masses usually
does not improve the quality of fit compared to the decoupling limit,
for fixed values of other quantities.
Although not statistically significant, the
precision electroweak data may have a  mild preference for
decoupled scalars and light gauginos/higgsinos over any
other form of supersymmetry breaking patterns. 

In short, the precision electroweak data is compatible
with split supersymmetry spectrum for all values of gaugino
and higgsino masses above direct collider limits.  Near the
direct limits, the overall fit improves by nearly a full
unit in $\Delta\chi^2$.
A priori, fitting to precision electroweak
observables did not have to be favorable to split supersymmetry,
and could have been incompatible for light gauginos and/or higgsinos.
As it is, the improved fits are mildly encouraging for the scenario.

{\bf Note Added:} Following the appearance of the present work, a recent
interesting paper~\cite{Marandella:2005wc} suggests that high-energy 
LEP2 data can
contribute substantively to precision electroweak analysis of light
superpartners.  Although we have not independently confirmed this
analysis, we want to bring it to the reader's attention. If correct, the
conclusion from doing a precision electroweak analysis with that super-set
of observables would eliminate the mild preference for light charginos and
neutralinos. As noted in ref.~\cite{Marandella:2005wc}, 
the analysis might change
again after all $e^+ e^- \rightarrow e^+e^-$ data above the $Z$ pole
becomes available.

\addcontentsline{toc}{section}{Appendix: Useful functions}
\section*{Appendix: Useful functions}
\label{appendixA} \renewcommand{\theequation}{A.\arabic{equation}}
\setcounter{equation}{0}
\setcounter{footnote}{1}

The kinematic loop-integral functions needed above are given in terms of
\beq
B(x,y) &=& -\int_0^1 dt \>\lnbar [t x + (1-t) y - t (1-t) s - i \epsilon]
\eeq
where $s=p^2$ is the external momentum invariant in a ($+$$-$$-$$-$) metric,
and
\beq
\lnbar X \equiv \ln(X/Q^2)
\eeq 
where $Q$ is the renormalization scale.
In the text, the arguments $(x,y)$ are particle names and should be 
interpreted to substitute $x\to m^2_x$ and $y\to m^2_y$ into the 
equations.

The other functions used in the text are defined in terms of $B(x,y)$:
\beq
H(x,y) &=& [2s-x-y-(x-y)^2/s] B(x,y)/3 
\nonumber \\ &&
+ 
[2 x \lnbar x + 2 y \lnbar y - 2s/3 + 
(y-x)(x \lnbar x - x - y \lnbar y +y)/s]/3
,
\\
F(x,y) &=& H(x,y) + (x+y-s) B(x,y),
\\
G(x) &=& H(x,x) + 2 x B(x,x) .
\eeq
These can be expanded in powers of $s$ according to:
\beq
B(x,y) &=& b_0(x,y) + s b_1(x,y) + s^2 b_2(x,y) + \ldots\\
H(x,y) &=& h_0(x,y) + s h_1(x,y) + s^2 h_2(x,y) + \ldots\\
F(x,y) &=& f_0(x,y) + s f_1 (x,y) + s^2 f_2(x,y) + \ldots\\
G(x) &=& s g_1 (x) + s^2 g_2(x) + \ldots
\eeq
with coefficients that follow from
\beq
b_0(x,y) &=& [x - x \lnbar x -y + y\lnbar y ]/(x-y) 
,
\\
b_1(x,y) &=& [2 x y \ln (y/x) + x^2 - y^2]/2(x-y)^3
,
\\
b_2(x,y) &=& [6 x y (x+y) \ln (y/x) + x^3 - y^3 + 9 x^2 y - 9 x y^2]/6 (x-y)^5
,
\\
b_3(x,y) &=& [12 x y (x^2 + 3 x y + y^2) \ln(y/x) + x^4 - y^4 + 
28 x^3 y - 28 x y^3]/12 (x-y)^7,
\phantom{xxxxx}
\\
b_4 (x,y) &=& 
[60 x y (x^3 + 6 x^2 y + 6 x y^2 + y^3) \ln(y/x)
+ 3 x^5 - 3 y^5 + 175 x y (x^3 - y^3) 
\nonumber \\ && 
+ 300 x^2 y^2 (x-y)]/60(x-y)^9
\eeq
and
\beq
&& b_0(x,x) = -\lnbar(x),\qquad
b_1(x,x) = 1/6x,\qquad
b_2(x,x) = 1/60x^2,
\phantom{xxxx}
\\
&&
b_3(x,x) = 1/420 x^3,
\qquad
b_4(x,x) = 1/2520 x^4
.
\eeq
The expansions converge provided that $\sqrt{s} < \sqrt{x}+ \sqrt{y}$
(in other words, below the threshold branch cut).

\section*{Acknowledgements}
The work of SPM was  supported by the National Science Foundation under 
Grant No.~PHY-0140129.  KT acknowledges support from the National Science
Foundation.  The work of JDW was supported
in part by the Department of Energy and the Michigan Center for
Theoretical Physics. We thank G. Kane and T. Wang for discussions.

\end{document}